\documentclass[aps,showpacs,nofootinbib,twocolumn]{revtex4}

\usepackage{graphicx}
\usepackage{bm}
\usepackage{amsmath}
\usepackage{amssymb}
\usepackage{hyperref}
\usepackage{epsfig}

\newcommand{\xbj}{x}
\newcommand{\zh}{z_h}
\newcommand{\nslash}{\kern 0.2 em n\kern -0.50em /}
\newcommand{\kslash}{\kern 0.2 em k\kern -0.45em /}
\newcommand{\lslash}{\kern 0.2 em l\kern -0.50em /}
\newcommand{\pslash}{\kern 0.2 em p\kern -0.50em /}
\newcommand{\Sslash}{\kern 0.2 em S\kern -0.50em /}
\newcommand{\Pslash}{\kern 0.2 em P\kern -0.50em /}
\newcommand{\Dslash}{\kern 0.2 em D\kern -0.65em /\kern 0.15em}

\newcommand{\bp}{\boldsymbol{p}_T}
\newcommand{\bP}{\boldsymbol{P}_T}
\newcommand{\bk}{\boldsymbol{k}_T}

\newcommand{\eps}{\epsilon}
\newcommand{\ssh}{\!\!\!/}

\newcommand{\Tr}{\operatorname*{Tr}\nolimits}

\newcommand{\ph}{\phi_h}

\begin{document}

\title{Beam single spin asymmetry of neutral pion production in semi-inclusive deep inelastic scattering}

\author{Wenjuan Mao}
\author{Zhun Lu}\email{zhunlu@seu.edu.cn}\affiliation{Department of Physics, Southeast University, Nanjing
211189, China}

\begin{abstract}
We study the beam spin asymmetry $A_{LU}^{\sin\phi_h}$ in semi-inclusive $\pi^0$ electroproduction contributed by the $T$-odd twist-3 distribution function $g^\perp(x,k_T^2)$.
We calculate this transverse momentum dependent distribution function for the $u$ and $d$ quarks inside the proton in a spectator model including the scalar and the axial-vector diquark components.
Using the model results, we estimate the asymmetry $A_{LU}^{\sin\phi_h}$ in the $e p \rightarrow e^\prime \pi^0 X$ process in which the lepton beam is longitudinally polarized.
The model prediction is compared with the data measured by the CLAS and HERMES Collaborations, and it is found that our numerical results agree with the experimental data reasonably.
Especially, our results can well describe the CLAS data at the region where the Bjorken $x$ and the pion transverse momentum is not large.
We also make a prediction on the asymmetry $A_{LU}^{\sin\phi_h}$ in $\pi^0$ electroproduction at CLAS12 using the same model calculation.
\end{abstract}


\maketitle

\section{Introduction}

Understanding the origins of the single spin asymmetries (SSAs) appearing in high-energy semi-inclusive processes is one of the important goals of QCD spin physics~\cite{bdr,D'Alesio:2007jt,Barone:2010ef,Boer:2011fh}.
Substantial SSAs have been measured by the HERMES Collaboration~\cite{hermes00,hermes01,hermes05,hermes07,hermes09,hermes10}
, the COMPASS Collaboration~\cite{compass05,compass06,compass10,Adolph:2012sn,Adolph:2012sp}, and the Jefferson Lab (JLab)~\cite{clas04,clas10,Qian:2011py,Aghasyan:2011ha,Aghasyan:2011zz} in semi-inclusive deep inelastic scattering(SIDIS).
It is found that the $T$-odd transverse momentum dependent (TMD) parton distribution functions (DFs)~\cite{Sivers:1991prd,Anselmino:1998plb,Boer:1998prd,
Brodsky:2002plb,Collins:2002plb,jy02} or fragmentation functions (FFs)~\cite{Collins:1993npb}, under the TMD factorization~\cite{Ji:2004wu} framework, play central roles in the observed SSAs.
The $T$-odd TMDs describe the correlations between the transverse motion of the parton and its own spin or the spin of the initial-state hadron,
thereby encoding much richer information about the partonic structure as well as the QCD dynamics inside hadrons than what can be learned from the collinear DFs.

In the particular case of a longitudinally polarized beam colliding on an unpolarized target,
an asymmetry with $\sin\phi_h$ modulation, the so-called beam SSA, emerges.
The CLAS Collaboration~\cite{clas04,Aghasyan:2011ha,Aghasyan:2011zz} at JLab and the HERMES Collaboration~\cite{hermes07} have measured this asymmetry in pion electroproduction in the magnitude of
several percents that cannot be explained by perturbative QCD~\cite{Ahmed:1999ix}.
Different mechanisms have been proposed to generate such asymmetry, such as the Boer-Mulders effect~\cite{Yuan:2004plb} and the Collins effect~\cite{Efremov:2002ut,Gamberg:2003pz}, involving the chiral-odd distribution or fragmentation functions.
In Refs.~\cite{Metz:2004epja,Afanasev:2003ze,Bacchetta:g2004plb}, a new source contributing to the beam SSA has been identified, either from the model calculations~\cite{Metz:2004epja,Afanasev:2003ze} or from the updated decomposition of the unpolarized quark-quark correlator~\cite{Bacchetta:g2004plb}, where the twist-3 TMD  $g^\perp(x,\bm k_T^2)$ plays a crucial role.
As a $T$-odd chiral-even TMD, $g^\perp$ can be regarded as an analog of the Sivers function~\cite{Sivers:1991prd} at the twist-3 level, because both of them require quark transverse motion as well as initial- or final-state interactions~\cite{Brodsky:2002plb,Collins:2002plb,jy02} via soft-gluon exchanges to receive nonzero contributions.
Therefore, studying beam SSAs may provide a unique opportunity to unravel the role of quark spin-orbit correlation at twist 3.

In this work, we present an analysis on the beam SSA $A_{LU}^{\sin\phi_h}$ in neutral pion production, based on the effect from $g^\perp$.
We calculate the function $g^\perp(x,\bm k_T^2)$ for the $u$ and $d$
quarks inside the proton, using the spectator model with scalar and
axial-vector diquarks.
Different types of spectator model have been widely used to calculate TMDs for the nucleon~\cite{Jakob:1997npa,Brodsky:2002plb, Boer:2002ju,Goldstein:2002vv, Gamberg:2003ey,Bacchetta:2004plb,Gamberg:2007wm,Meissner:2007rx,
Bacchetta:2008prd,Bacchetta:2010si} and the pion~\cite{Lu:2004hu,Lu05,Meissner:2008ay,Gamberg:2009uk}.
We will adopt a specific model given in Ref.~\cite{Bacchetta:2008prd}, in which
the authors consider the isospin of vector diquarks to distinguish the
isoscalar ($ud$-like) and isovector ($uu$-like) spectators.
Furthermore, in that model the free parameters are fixed by reproducing the parametrization of unpolarized and longitudinally polarized parton distributions.
The above-mentioned feature of the model allows us to perform the phenomenological analysis in a deep sense.
Using the calculated $g^{\perp u}$ and $g^{\perp d}$, we estimate the beam SSA $A_{LU}^{\sin\phi_h}$ in neutral pion production at the kinematics of CLAS and compare our results with the CLAS data~\cite{Aghasyan:2011ha} measured recently with high precision. We also make a comparison between our calculation and the $\pi^0$ data measured by the HERMES Collaboration~\cite{hermes07} for further testing. Finally, we present the prediction of $A_{LU}^{\sin\phi_h}$ for $\pi^0$ production at CLAS12.

The paper is organized as follows.
In Sec.~\ref{formulation}, we present the details on the calculation of $g^\perp$ in the spectator model with an axial-vector diquark, and discuss the flavor, $x$ and $k_T$ dependencies of $g^\perp$.
In Sec.~\ref{BSAs}, we analyze the beam SSA $A_{LU}^{\sin\phi_h}$  in $\pi^0$ production numerically at CLAS, HERMES, and CLAS12 based on the calculated $g^\perp$ for the $u$ and $d$ quarks.
We address our conclusion in Sec.~\ref{conclusion}.

\section{$g^\perp$ of the proton in the spectator model with an axial-vector diquark}
\label{formulation}

In this section, we present the detailed calculation on the $T$-odd twist-3 TMD $g^\perp(x,\bm k_T^2)$ of the proton in a spectator model with axial-vector diquark. The starting point of the calculation is the gauge-invariant quark-quark correlator for the unpolarized nucleon
\begin{align}
\Phi^{[+]}(x,k_T)=&\int {d\xi^- d^2\xi_T\over (2\pi)^3}e^{ik\cdot\xi}
\langle P|\bar{\psi}_j(0)\mathcal{L}[0^-,\infty^-]\nonumber\\
&\times\mathcal{L}[0_T,\xi_T]\mathcal{L}[\infty^-,\xi^-]\psi_i(\xi)|P\rangle\,,
\end{align}
where $[+]$ denotes that the gauge link appearing in $\Phi$ is future-pointing, corresponding to the SIDIS process.
As an interaction-independent twist-3 distribution, $g^{\perp}$ naturally appears in the decomposition of the quark-quark correlator at the subleading order of $1/P^+$ expansion, after including the light-cone vector $n_-$ that defines the direction along which the path-order exponential runs~\cite{Bacchetta:g2004plb,Goeke:2005plb}
\begin{align}
\Phi^{[+]}(x,k_T)\bigg{|}_{\mathcal{O}\left({M\over P^+}\right)} &= \int d k^- \Phi^{[+]}(P,k;n_-) \nonumber\\
&=  {M\over 2P^+}\left\{g^\perp\gamma_5{\epsilon_T^{\rho\sigma}\gamma_\rho k_{T\sigma}\over M}+\cdots\right\},
\end{align}
here, $\cdots$ stands for the other twist-3 TMDs that
are not taken into account in this paper.
The distribution $g^\perp$, therefore, can be deduced from $\Phi^{[+]}(x,k_T)$ by taking the trace with proper Dirac matrices
\begin{align}
- \frac{\eps_{T}^{\alpha\rho} k_{T \rho}^{}}{P^+} \,
  g^{\perp}(x,\bm{k}_{T}^{2})=\frac{1}{2}\Tr[\Phi^{[+]}\gamma^{\alpha}\gamma_5]. \label{phitr}
  \end{align}

Since $g^\perp$ is a $T$-odd DF, we need to consider the effect of the gauge link to generate a nonzero result, in analogy to the Sivers function and the Boer-Mulders function.
However, calculations on $g^\perp$ based on the scalar diquark model (with pointlike proton-quark-diquark coupling) as well as the quark-target model~\cite{Gamberg:2006plb} show that $g^\perp$ and the other twist-3 $T$-odd TMDs suffer light-cone divergence.
This feature is in contrast to the case of leading-twist $T$-odd TMDs and has been recognized as a theoretical challenge in deriving a TMD factorization proof in SIDIS at twist 3 \cite{Gamberg:2006plb,Bacchetta:2008jhep}.
In Refs.~\cite{Gamberg:2006plb,Gamberg:2008yt}, the authors pointed out that light-cone divergence can be avoided by means of a phenomenological
approach in which form factors are applied for the proton-quark-diquark coupling.
In Ref.~\cite{Lu:2012plb}, all sixteen twist-3 $T$-odd TMDs were calculated in the scalar diquark model by adopting the dipolar form factor for the proton-quark-diquark coupling. In this paper, we extend the calculations in Refs.~\cite{Gamberg:2006plb,Lu:2012plb} using the spectator model with an axial-vector diquark to obtain $g^\perp$ for both the $u$ and $d$ quarks.
The model we apply in the calculation is the version developed in Ref.~\cite{Bacchetta:2008prd}, which was originally proposed in Ref.~\cite{Jakob:1997npa}.
The same model has been adopted to calculate various twist-3 quark-gluon-quark correlators in Ref.~\cite{Kang:2010hg}.
The common feature of spectator models is that the proton with mass $M$ is supposed to be constituted by a quark with mass $m$ and a diquark with mass $M_X$, and the diquark $X$ can be either a scalar one (denoted by $s$) or an axial-vector one (denoted by $v$).

We perform the calculation in the Feynman gauge and expand the gauge link to the first order (one-gluon exchange) to obtain the quark-quark correlators contributed by the scalar diquark and the axial-vector diquark components
\begin{widetext}
\begin{align}
\begin{split}
\Phi_{sij}(x,k_T)&\equiv -i e_q  {1\over(2\pi)^4}\int {d^4q\over (2\pi)^4} \int d k^{-}
  {1\over q^+ +i\epsilon}
\left[\bar{U}(P,S)\Upsilon_s(k^2)
 { (k\ssh +m)\over k^2-m^2+i\epsilon}\right]_j
 \left[
 {  1 \over (P-k)^2-M_s^2+i\epsilon} \right.
\\
& \times \left.    {\Gamma_s^{+}\over q^2 +i\varepsilon}
 {1 \over (P-k+q)^2-M_s^2+i\epsilon}\right]
 \left[
 { (k\ssh-q\ssh+m)\over (k-q)^2-m^2+i\epsilon}\Upsilon_s((k-q)^2) U(P,S)\right]_i
 \bigg|_{
k^{+} =xP^+ } +H.c.,
\end{split}
\label{phis1}\\
\begin{split}
\Phi_{vij}(x,k_T)&\equiv -i e_q  {1\over (2\pi)^4}\int {d^4q\over (2\pi)^4} \int d k^{-}
  {1\over q^+ +i\epsilon}
\left[\bar{U}(P,S)\Upsilon_v^{\rho}(k^2)
 { (k\ssh +m)\over k^2-m^2+i\epsilon}\right]_j
 \left[
 {  d_{\rho\alpha}(P-k) \over (P-k)^2-M_v^2+i\epsilon} \right.
\\
& \times\left.  {\Gamma^{+,\alpha\beta}_v\over q^2 +i\varepsilon}
 {d_{\beta\sigma}(P-k+q) \over (P-k+q)^2-M_v^2+i\epsilon}\right]
 \left[
 { (k\ssh-q\ssh+m)\over (k-q)^2-m^2+i\epsilon}\Upsilon_v^\sigma((k-q)^2) U(P,S)\right]_i
 \bigg|_{
k^{+} =xP^+} +H.c.,
\end{split}
\label{phiv1}
\end{align}
\end{widetext}
 where $e_q$ is the charge of the quark, $\Upsilon_{s/v}$ denotes the nucleon-quark-diquark vertex with the form~\cite{Jakob:1997npa}
\begin{align}
\Upsilon_s(k^2) = g_s(k^2),~~
\Upsilon_v^\mu(k^2)={g_v(k^2)\over \sqrt{2}}\gamma^\mu\gamma^5,
\end{align}
and $\Gamma_s^\mu $ or $\Gamma_v^{\mu,\alpha\beta}$ is the vertex between the gluon and the scalar diquark or the axial-vector diquark
\begin{align}
 \Gamma_s^\mu  =& ie_s (2P-2k+q)^\mu, \\
 \Gamma_v^{\mu,\alpha\beta} =&  -i e_v [(2P-2k+q)^\mu g^{\alpha\beta}-(P-k+q)^{\alpha}g^{\mu\beta}\nonumber\\
 &-(P-k)^\beta g^{\mu\alpha}],
 \end{align}
with $e_{s/v}$ denoting the charge of the scalar/axial-vector diquark.
In Eq.(\ref{phiv1}) we use $d_{\mu\nu}$ to denote the summation over
the polarizations of the axial-vector diquark for which we choose
the following form~\cite{Brodsky:2000ii}
\begin{align}
 d_{\mu\nu}(P-k)  =& \,-g_{\mu\nu}\,+\, {(P-k)_\mu n_{-\nu}
 \,+ \,(P-k)_\nu n_{-\mu}\over(P-k)\cdot n_-}
\nonumber\\& - \,{M_v^2 \over\left[(P-k)\cdot n_-\right]^2 }\,n_{-\mu} n_{-\nu} .\label{d1}
 \end{align}
The advantage of the above choice has been argued
in Ref.~\cite{Bacchetta:2008prd}.
To obtain finite results for $g^\perp$, we also choose
the dipolar form factor for $g_X(k^2)$
\begin{align}
g_X(k^2)&= N_X {k^2-m^2\over |k^2-\Lambda_X^2|^2} \nonumber \\
&= N_X{(k^2-m^2)(1-x)^2\over
(\bm k_T^2+L_X^2)^2},
 \end{align}
for $X=s,\,v$, where $\Lambda_X$ is the cutoff parameter, $N_X$ is the coupling constant (which also serves as the normalization constant), and
$L_X^2$ has the form
\begin{align}
L_X^2=(1-x)\Lambda_{X}^2 +x M_{X}^2-x(1-x)M^2.
\end{align}

Performing the integrations in Eqs.~(\ref{phis1}) and (\ref{phiv1}) over $k^-$, $q^+$ and $q^-$, the quark-quark correlators for the unpolarized nucleon in the spectator model are simplified as
\begin{widetext}
\begin{align}
  \Phi_s
(x,k_T)
&\equiv
-i e_qe_s N_{s}^2  { (1-x)^3\over 32\pi^3 P^+}{1\over (L_{s}^2+\bm{k}_T^2)^2}\int {d^2 \bm q_T\over (2\pi)^2}
{ \left[(\kslash -q\ssh+m) (\Pslash+M)(\kslash +m)\right]
\over \bm q_{T}^2  (L_s^2+(\bm{k}_T-\bm{q}_T)^2)^2}
 \bigg|_{\begin{subarray}{l} q^+ = 0  \\ k^{+} =xP^+  \end{subarray}},\\
 \Phi_{v}
(x,k_T)
&\equiv
-i e_q N_v^2  { (1-x)^2\over 128\pi^3 (P^+)^2}{1\over (L_v^2+\bm{k}_T^2)^2}\int {d^2 \bm q_T\over (2\pi)^2} \,
 d_{\rho\alpha}(P-k)\, (-i\Gamma^{+,\alpha\beta}) \, d_{\sigma\beta}(P-k+q) \nonumber\\
&\times{ \left[(\kslash -q\ssh+m) \gamma^\sigma(\Pslash-M)\gamma^\rho (\kslash +m)\right]
\over \bm q_T^2  (L_v^2+(\bm{k}_T-\bm{q}_T)^2)^2}
 \bigg|_{\begin{subarray}{l}
 q^+ = 0  \\
k^{+} =xP^+  \end{subarray}}.
\end{align}
\end{widetext}
Using Eq.~(\ref{phitr}) and performing the integration
over $\bm q_T$, we arrive at the
expressions for the distribution $g^\perp$ from the scalar and the axial-vector diquark components~\footnote{There is a typo in Ref.~\cite{Lu:2012plb}, the factor
$(1-x)\Lambda^2_s+(1+x)M_s^2 -(1-x)(1-2x)M^2$ in the numerators of the right-hand side of Eqs.~(32) and (36)  should be
$(1-x)\Lambda^2_s+(1+x)M_s^2 -(1-x)M^2$.}
\begin{widetext}
\begin{align}
g^{\perp s}(x,\bm k_{T}^2)
 &=-{ N_s^2(1-x)^2\over(32\pi^3)}{e_se_q\over4\pi} \left[\frac{(1-x)\Lambda^2_s+(1+x)M_s^2 -(1-x)M^2}{
 {L_s^2 (L_s^2 +\bm k_T^2)^3}}\right],\\
g^{\perp v}(x,\bm k_{T}^2) &= {N_v^2(1-x)^2 \over(32\pi^3)}{e_ve_q\over4\pi}\left[{(1-x)(xM+m)^2+(1-x)^2 M^2-M_v^2+xL_v^2\over(1-x)L_v^2(L_v^2+\bm k_T^2)^3} \right.\nonumber\\
&\left.-
{x\over(1-x) \bm k_T^2(L_v^2+\bm k_T^2)^2}\ln\left({L_v^2+\bm k_T^2\over L_v^2}\right)\right].
\end{align}
\end{widetext}

The function $g^\perp$ for the $u$ and $d$ quarks can be constructed by $g^{\perp s}$ and $g^{\perp v}$.
Here, we follow the approach in Ref.~\cite{Bacchetta:2008prd} in which
the two isospin states of the vector diquark are distinguished;
that is, the vector isoscalar diquark $v(ud)$ is denoted by $a$
with mass $M_a$ and cutoff parameter $\Lambda_a$,
while the vector isovector diquark $v(uu)$ is denoted by
$a^\prime$ with different mass $M_{a^{\prime}}$ and cutoff parameter
$\Lambda_{a^\prime}$.
Hence, by applying the relation between quark flavors and diquark types, we can obtain
the distributions  $g^{\perp u}$ and $g^{\perp d}$
by the following form~\cite{Bacchetta:2008prd}:
\begin{align}
g^{\perp u}&=c_s^2 \,g^{\perp s} + c_a^2\,g^{\perp a}, \\
g^{\perp d}&=c_a^{\prime 2} \,g^{\perp a^\prime},
\end{align}
where $c_s$, $c_a$, and $c_{a^\prime}$ represent different
couplings that are the free parameters of the model.
The same form also holds for the other TMDs.

For the parameters $M_X$,  $\Lambda_X$ and $c_X$ ($X=s,a, a^\prime)$ needed in the calculation, we also adopt the values from Ref.~\cite{Bacchetta:2008prd},
as shown in the first three columns of Table.~\ref{Tab.1}.
\begin{table}
\begin{tabular}{|c||c|c|c|c|}
\hline
Diquark & $M_X$(GeV) & $\Lambda_X$(GeV) & $c_X$  & $N_X$ \\
\hline
Scalar $s$ $(ud)$ & 0.822 & 0.609  &~ 0.847 ~& 11.400 \\
\hline
Axial-vector $a$ $(ud)$ & 1.492  & 0.716  & 1.061  & 28.277 \\
\hline
Axial-vector $a'$ $(uu)$ & 0.890  & 0.376  & 0.880 & 4.091 \\
\hline
\end{tabular}
\caption{Values for the free parameters of the model taken from Ref.~\cite{Bacchetta:2008prd}, which are fixed by reproducing the parametrization of unpolarized~\cite{zeus} and longitudinally polarized~\cite{grsv01} parton distributions.}
\label{Tab.1}
\end{table}
The fourth column shows the values for the corresponding normalization constants $N_X$ ($X=s,a, a^\prime)$, which are obtained from the normalization condition
for the unpolarized TMD. The quark mass is chosen as $m=0.3\,\text{GeV}$.
To convert our calculation to real QCD, we use the following replacement for the combination of the charges of the quark $q$ and the spectator diquark $X$
\begin{align}
{e_qe_X\over 4 \pi}\rightarrow -  C_F \alpha_s
\end{align}
and choose $\alpha_s \approx 0.3 $ in our calculation. The minus sign in the above equation comes from the fact that in QCD, a hadron is color-neutral; thus, the color charges $e_q$ and $e_X$ should have the opposite signs.

\begin{figure}
  \includegraphics[width=0.49\columnwidth]{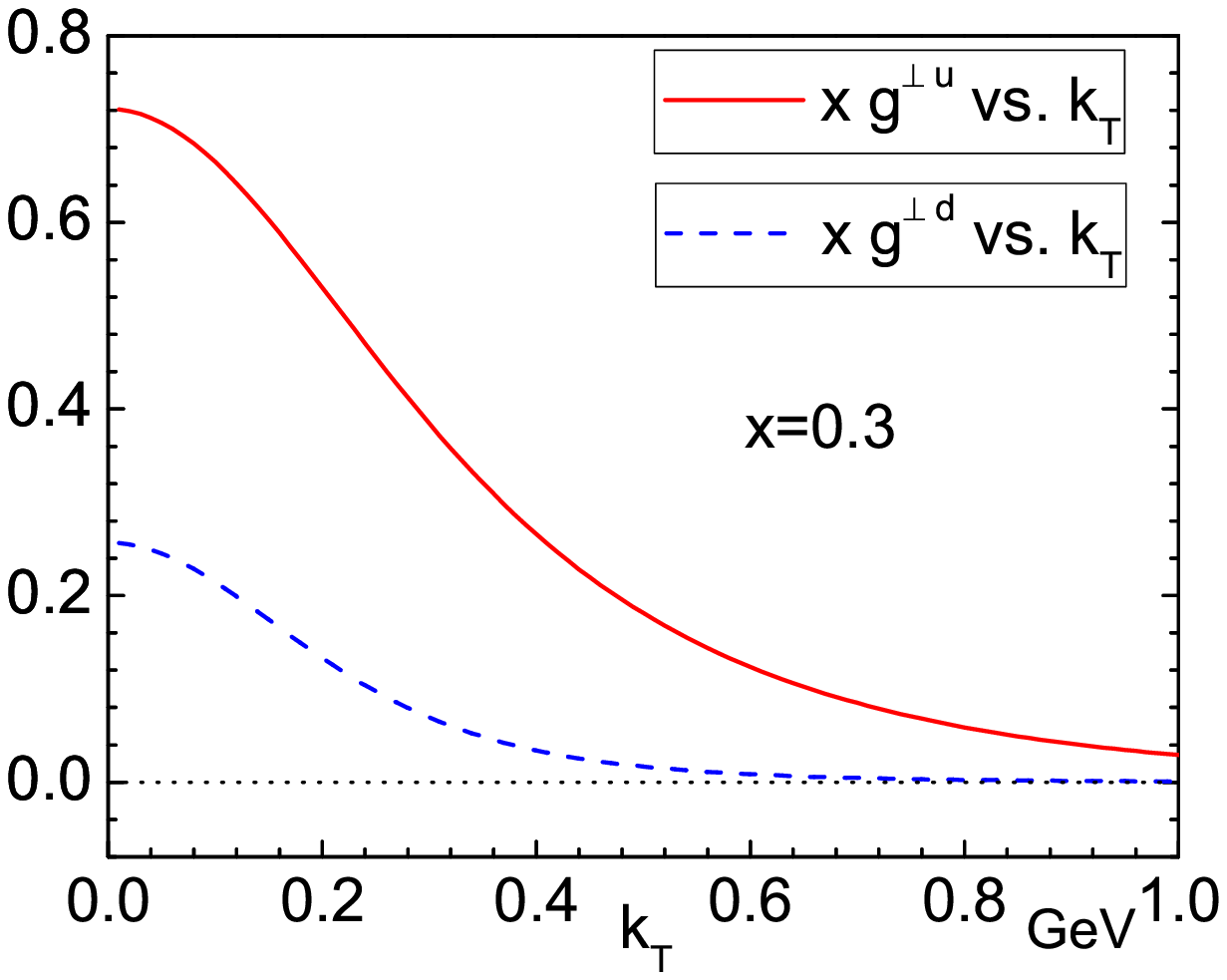}
  \includegraphics[width=0.49\columnwidth]{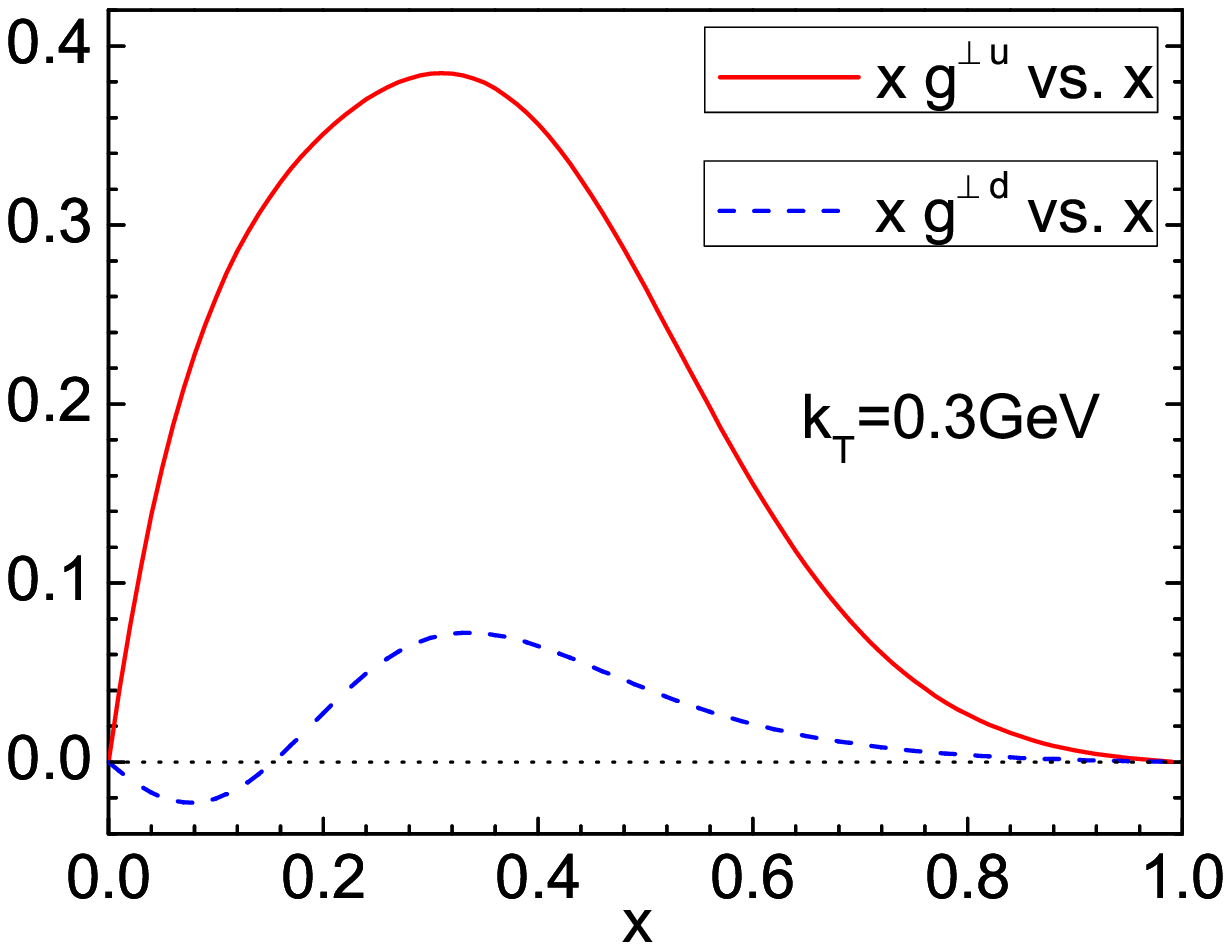}\\
  \caption{Left panel: model results for $x g^{\perp u}$ (solid line) and $x g^{\perp d}$ (dashed line) as  functions of $k_T$ at $x=0.3$; right panel: model results for $x g^{\perp u}$ (solid line) and $x g^{\perp d}$ (dashed line)  as functions of $x$ at  $k_T=0.3\,\text{GeV}$ .}\label{FIG.1}
\end{figure}

In the left panel of Fig.~\ref{FIG.1}, we plot the functions $x g^{\perp u}$ (solid line)
and $x g^{\perp d}$ (dashed line) vs $k_T$ at $x=0.3$, while in the right panel of the same figure, we display the $x$ dependence of $x g^{\perp u}$
and $x g^{\perp d}$ at $k_T=0.3 \,\text{GeV}$.
The plots in Fig.~\ref{FIG.1} correspond to the results in the scale $\mu_0^2 = 0.3\,\text{GeV}^2$, which is the scale used in Ref.~\cite{Bacchetta:2008prd} to fit the parametrization of $f_1(x)$~\cite{zeus}.
As we can see from Fig.~\ref{FIG.1}, the dominance of $u$ quark contribution is evident in the adopted spectator model; that is, $g^{\perp u}$ is several times larger than $g^{\perp d}$ by size.
Our results show that $g^{\perp u} $ is positive for all $x$ and $k_T$ regions; while $g^{\perp d}$ is negative in the small $x$ region and turns to be positive in the region $x>0.15$, i.e., there is a node in the $x$ dependence of $g^{\perp d} $.
Also, the $k_T$ dependencies of the $u$ and $d$ quark distributions are different since $g^{\perp d}$ approaches zero faster than $g^{\perp u}$ when
$k_T$ increases.

\section{numerical results for beam spin asymmetry}
\label{BSAs}

In this section, we will use our model resulting $g^\perp$ to calculate
the beam SSA $A_{LU}^{\sin\phi_h}$ in  $\pi^0$ electroproduction,
as precise measurements on $A_{LU}^{\sin\phi_h}$ of a neutral pion for different $x$ and $P_T$ bins have been performed by the CLAS Collaboration~\cite{Aghasyan:2011ha} at JLab recently, in SIDIS by a $5.776\,\text{GeV}$ longitudinally polarized electron beam off an unpolarized hydrogen target.
Earlier, the HERMES Collaboration measured the beam SSAs for neutral and charged pions using a $27.6\,\text{GeV}$ beam.
We compare the results for $A_{LU}^{\sin\phi_h}$ with the neutral pion data from CLAS and HERMES to test our model calculation.
Then, we will make new prediction on the asymmetry $A_{LU}^{\sin\phi_h}$  for $\pi^0$ electroproduction at CLAS12 using the same model calculation to study the prospects to access $g^\perp$ at CLAS after the $12 \,\text{GeV}$ upgrade is realized.

The semi-inclusive leptoproduction process that we study can be expressed as
\begin{align}
e (\ell) \, + \, p (P) \, \rightarrow \, e' (\ell')
\, + \, h (P_h) \, + \, X (P_X)\,,
\label{sidis}
\end{align}
where $\ell$ and $\ell'$ are the four-momentum of the incoming and scattered lepton, and $P$ and $P_h$ are the four-momentum of the target nucleon and the detected final-state hadron $h$, respectively.

The variables to express the SIDIS cross section are defined as
\begin{align}
&x = \frac{Q^2}{2\,P\cdot q},~~~
y = \frac{P \cdot q}{P \cdot l},~~~
z = \frac{P \cdot P_h}{P\cdot q},~~~\gamma={2M x\over Q},~~~\nonumber\\
&Q^2=-q^2, ~~~
s=(P+\ell)^2,~~~
W^2=(P+q)^2,~~~
\end{align}
where $q=\ell-\ell'$ is the four-momentum of the virtual photon and $W$ is the invariant mass of the hadronic final state.

\begin{figure}
  \includegraphics[width=0.9\columnwidth]{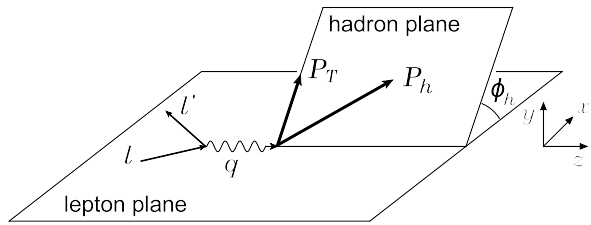}
 \caption{The kinematical configuration for the SIDIS process. The initial and scattered leptonic momenta define the lepton plane ($x-z$ plane), while the detected hadron momentum together with the $z$ axis identify the hadron production plane.}
 \label{sidiskine}
\end{figure}

The reference frame we adopt here is that the virtual photon and the target proton are collinear and along the $z$ axis, with the photon moving toward the target in the positive $z$ direction, as shown in Fig.~\ref{sidiskine}. We use $\bk$ to denote the intrinsic transverse momentum of the quark inside the proton for the DFs, with $\bP$ to denote the transverse momentum of the detected hadron. The transverse momentum of the hadron $h$ with respect to the direction of the fragmenting quark is denoted by $\bp$, which appears in the TMD FFs. The azimuthal angle between the lepton and the hadron planes is defined as $\ph$, following the Trento convention \cite{Bacchetta:2004prd}.

Up to subleading order of $1/Q$, the differential cross section of SIDIS for a longitudinally polarized beam (with helicity $\lambda_e$) off an unpolarized hadron has the following general expression~\cite{Bacchetta:0611265}:
\begin{align}
\label{HLT}
\frac{d\sigma}{d\xbj dy\,d\zh dP^2_T d\ph} =&\frac{2\pi \alpha^2}{\xbj y Q^2}\frac{y^2}{2(1-\varepsilon)}
 \Bigl( 1+ \frac{\gamma^2}{2\xbj} \Bigr) \left\{ F_{UU}\right.
\nonumber\\& \left. + \lambda_e \sqrt{2\varepsilon(1-\varepsilon)} \sin \phi_h \,\,F^{\sin \ph}_{LU}\right\},
\end{align}
where $F_{UU}$ is the helicity-averaged structure function, while $F_{LU}^{\sin\phi_h}$ is helicity-dependent structure function resulting from the antisymmetric part of the unpolarized hadronic tensor.
The first and second
subscript of the above structure functions indicate the polarization of beam and
target, respectively.
It is $F_{LU}^{\sin\phi_h}$ that gives rise to the $\sin\phi_h$ beam SSA.
The ratio of the longitudinal and transverse photon flux is given by
\begin{align}
\varepsilon=\frac{1-y-\gamma^2y^2/4}{1-y+y^2/2+\gamma^2y^2/4}.
\end{align}

In the parton model, the two structure functions in Eq.~(\ref{HLT}) can be expressed as the convolution of TMD DFs and FFs, based on the tree-level factorization adopted in Ref.~\cite{Bacchetta:0611265}.
With the help of the notation
\begin{align}
\mathcal{C}[w fD] =& x\sum_q e_q^2\int d^2\bm k_T \int d^2 \bm p_T \delta^2(z\bm k_T-\bm P_T+\bm p_T)\nonumber\\
& \times w(\bm k_T, \bm p_T)f^q(x,\bm k_T^2) D^q(z,\bm p_T^2),
\end{align}
$F_{UU}$ and $F_{LU}^{\sin\phi_h}$ can be written as~\cite{Bacchetta:0611265}
\begin{align}
F_{UU} & = \mathcal{C}[f_1 D_1], \label{FUU}\\
F^{\sin \ph}_{LU} & = \frac{2M}{Q} \,
\mathcal{C}\,
   \left [ \frac{\boldsymbol{\hat{P}_{T}}\cdot
    \boldsymbol{k_T}}{M}\left(\frac{M_h}{M}\, h_1^{\perp} \frac{\tilde{E}}{z} +\xbj\, g^\perp D_1\right) \right.\nonumber\\
    &\left.-
    \frac{\boldsymbol{\hat{P}_{T}} \cdot \boldsymbol{p_T}}{M_h}
         \left(\frac{M_h}{M}\,f_1\, \frac{\tilde{G^\perp }}{z} + \xbj\, e H_1^{\perp} \right) \right ] ,\label{FLU}
\end{align}
where $\hat {\bm P}_T= {\bP\over P_T}$ with $P_T =|\bP|$,
$M$ and $M_h$ are the nucleon and hadron masses, respectively. The functions $\tilde{G^\perp}$ and  $\tilde{E}$ are the interaction-dependent twist-3 FFs that come from the quark-gluon-quark correlator for FFs. The former one is $T$-odd, while the later one is $T$-even. They can be connected to the interaction-independent twist-3 FFs $G^\perp$~\cite{Bacchetta:2004plb} and $E$~\cite{Jaffe:1992npb} by the following relations \cite{Bacchetta:0611265}:
\begin{align}
\frac{\tilde{G^\perp }}{z}=\frac{G^\perp }{z}-\frac{m}{M_h}H_1^{\perp},~~~~~~~~
\frac{\tilde{E}}{z}=\frac{E}{z}-\frac{m}{M_h}D_1.
\end{align}

The beam-spin asymmetry $A_{LU}^{\sin\phi}$ in single-pion production off an unpolarized target thus is expressed as
\begin{widetext}
\begin{align}
A_{LU}^{\sin\phi}(P_T) = {\int dx \int dy \int dz \;\frac{1}{x y Q^2}\frac{y^2}{2(1-\varepsilon)}
   \times \,
    \Bigl( 1+ \frac{\gamma^2}{2x} \Bigr) \sqrt{2\varepsilon(1-\varepsilon)} \;F_{LU}^{\sin\phi} \over
     \int dx \int dy \int dz \;\frac{1}{x y Q^2}\frac{y^2}{2(1-\varepsilon)}
   \times \,
    \Bigl( 1+ \frac{\gamma^2}{2x} \Bigr) \;F_{UU} } \label{asy}
\end{align}
\end{widetext}
for the $P_T$-dependent asymmetry. The $x$-dependent and the $z$-dependent asymmetries can be defined in a similar way.

The structure function $F^{\sin \ph}_{LU}$ receives various contributions
from the convolution of the twist-2 and twist-3 TMD DFs and FFs, as shown in Eq.~(\ref{FLU}).
The $h_1^{\perp} E$ term, instead of the $ h_1^{\perp}
\tilde{E} $ in Eq.~(\ref{FLU}), has been calculated in Ref.~\cite{Yuan:2004plb}.
The contribution from the $g^\perp D_1$ term has been studied in Refs.~\cite{Metz:2004epja,Afanasev:2003ze,Bacchetta:g2004plb,
Afanasev:2006gw}.
These two terms generate the asymmetry through the effects of the $T$-odd distribution functions, namely, the twist-2 Boer-Mulders function $h_1^\perp$ ~\cite{Boer:1998prd} and the twist-3 $g^\perp$.
Each distribution represents a specific spin-orbit correlation of the initial quark inside the nucleon.
The beam SSA of the $\pi^+$ meson has also been analyzed~\cite{Efremov:2002ut,Gamberg:2003pz} based on the Collins effect $e H_1^\perp$, where $H_1^{\perp}$ is the $T$-odd Collins FF~\cite{Collins:1993npb}, and
$e$ is the chiral-odd twist-3 DF~\cite{Jaffe:1992npb,Schweitzer:2003prd}.

In the following, we will calculate the beam SSA in semi-inclusive pion electroproduction contributed by the $g^\perp D_1$ term. To do this, we first
neglect the quark-gluon-quark correlators (often referred to as the Wandzura-Wilczek approximation~\cite{Wandzura:1977qf}) for FFs, which is equivalent to setting all the functions with a tilde to zero.
It is worthwhile to point out that a calculation from spectator model~\cite{Gamberg:2008yt} as well as a model-independent analysis~\cite{Metz2008prl} of the $T$-odd collinear quark-gluon-quark
correlators shows that the gluonic (partonic) pole contributions for FFs vanish.
The FF $\tilde{G}^\perp(x,\bm p_T^2)$ appears in the decomposition of the $T$-odd part of the TMD quark-gluon-quark correlator~\cite{Boer:2003cm,Bacchetta:0611265}, for which the gluonic pole contribution should play an essential role.
Whether the vanishing gluonic pole matrix elements for collinear FFs can be generalized to the case of TMD FFs deserves further study~\cite{Gamberg:2010uw}.
Nevertheless, we ignore the $\tilde{G}^\perp$ and $\tilde{E}$ contributions based on the Wandzura-Wilczek approximation.
Therefore, there are only two terms inside the square brackets in the right-hand side of Eq.~(\ref{FLU}) remaining.
Moreover, we consider merely the beam SSA of the $\pi^0$ production, since
the fascinating fact that the favored and the unfavored Collins functions have similar sizes but opposite signs~\cite{Efremov:2006prd,Anselmino:2008jk} suggests that the $e H_1^\perp$ term leads vanishing beam SSA in $\pi^0$ electroproduction.
Specifically, the isospin symmetry determines that the $\pi^0$ FF should be the average of $\pi^+$ and $\pi^-$ FFs, so that $H_1^{\perp \pi^0/q} = (H_1^{\perp fav}+ H_1^{\perp unf})/2\approx 0 $.
Thus, in the following calculation, we can just take into account the term $g^\perp D_1$ and obtain
\begin{align}
\label{FLUa}
   F^{\sin \ph}_{LU}\approx &  \frac{2Mx}{Q} \,
  \sum_{q=u,d} e_q^2 \int d^2 \! \bk \,
   \biggl\{ \frac{\hat{\bm P}_T\cdot\bk} {M}
         \left[x\, g^{\perp q}(x,\bk^2)\right.\nonumber\\
         &\left.\times D_1^{q}\left(z,(\bP-z\bk)^2\right)\right]  \biggr\}\,.
\end{align}

For the unpolarized TMDs $f_1^q(x,\bk^2)$, we adopt the results from the same spectator model calculation~\cite{Bacchetta:2008prd} for consistency.
For the unpolarized TMD FFs $D_1^q\left(z,\bp^2\right)$, we assume that their $p_T$ dependencies have a Gaussian form
\begin{align}
D_1^q\left(z,\bp^2\right)=D_1^q(z)\, \frac{1}{\pi \langle p_T^2\rangle}
\, e^{-\bm p_T^2/\langle p_T^2\rangle}\,;
\end{align}
here, $\langle p_T^2\rangle$ is the Gaussian width for $ p_T^2$, which is chosen as $\langle p_T^2\rangle=0.2$ Gev$^2$, following the fitted result in Ref.~\cite{Anselmino:2005prd}.
For the integrated FFs $D_1^q(z)$, we adopt the Kretzer parametrization~\cite{kretzer2000}.

To perform the numerical calculation on the asymmetry $A_{LU}^{\sin \phi_h}$ in $\pi^0$ production at CLAS, we adopt the following kinematical cuts~\cite{Aghasyan:2011ha}:
\begin{align}
&0.4<z<0.7,~~ Q^2>1\, \textrm{GeV}^2, ~~ W^2>4\,\textrm{GeV}^2,\nonumber\\
&P_T>0.05\,\textrm{GeV},~~M_x(e \pi^0) > 1.5 \,\textrm{GeV},
\end{align}
where $M_x(e \pi^0)$ is the missing-mass value for the $e\pi^0$ system. Finally, in our calculation, we consider two different cases concerning the kinematical constraints on the intrinsic transverse momentum of the initial quarks~\cite{Boglione:2011}.
The first case is that we do not impose any constraint for $k_T$ in the calculation; the second case is that we consider the following kinematical constraints that are derived in Ref.~\cite{Boglione:2011}:
\begin{equation}
 \begin{cases}
k_{T}^2\leq(2-x)(1-x)Q^2, ~~~\textrm{for}~~0< x< 1 
; \\
k_{T}^2\leq \frac{x(1-x)} {(1-2x)^2}\, Q^2, ~~~~~~~~~~~~\textrm{for}~~x< 0.5.
\end{cases}\label{constraints}
 \end{equation}
The above constraints fix the upper limit for the allowed $k_T$ range;
thus, they can modify certain azimuthal asymmetries substantially, such as the
the twist-3 Cahn effect analyzed in Ref.~\cite{Boglione:2011}.
In our calculation of $A_{LU}^{\sin\phi_h}$, which is also a twist-3 observable, we find again that the kinematical constraints (\ref{constraints}) modify the sizes of the asymmetry in certain kinematical regions.

\begin{figure}
  \includegraphics[width=\columnwidth]{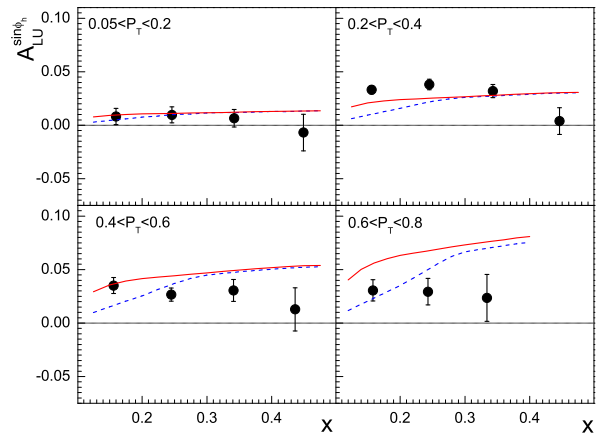}
  \caption{Beam SSA $A_{LU}^{\sin\phi_h}$ in $\pi^0$ electroproduction contributed by $g^\perp$ as a function of  $x$ for different $P_T$ ranges. The solid and the dashed lines correspond to the results without and with the kinematical constraints (\ref{constraints}) on $k_T$, respectively. Data are from Ref.~\cite{Aghasyan:2011ha}, the error bars for the data including the systematic and statistical uncertainties.  }\label{aluvx}
\end{figure}

\begin{figure}
  \includegraphics[width=\columnwidth]{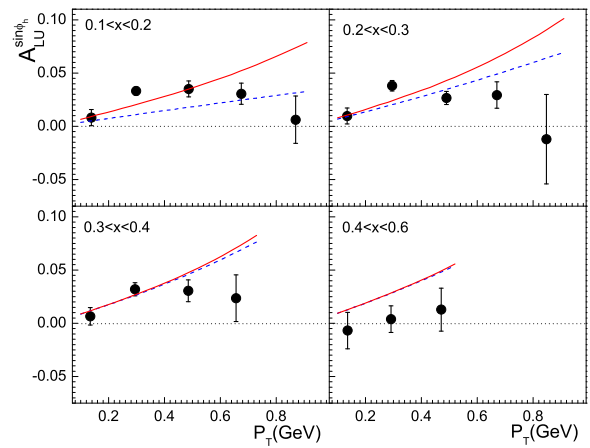}
  \caption{Beam SSA $A_{LU}^{\sin\phi_h}$ in $\pi^0$ electroproduction contributed by $g^\perp$ as a function of  $P_T$ for different $x$ ranges. The solid and the dashed lines correspond to the results without and with the kinematical constraints (\ref{constraints}) on $k_T$, respectively. Data are from Ref.~\cite{Aghasyan:2011ha}, the error bars for the data including the systematic and statistical uncertainties.  }\label{aluvpt}
\end{figure}

Figure~\ref{aluvx} shows the results of the beam SSA $A_{LU}^{\sin\phi_h}$ for $\pi^0$ production as a function of $x$, compared with the CLAS data (full circles) measured using a 5.776 GeV electron beam~\cite{Aghasyan:2011ha}.
The four panels correspond to the asymmetry integrated over four different $P_T$ ranges. From the comparison between the theoretical curves and the data, one can see that our results qualitatively describe the $x$ dependence of the asymmetry. Especially, for the ranges $0.05\,\textrm{GeV} <P_T <0.2\,\textrm{GeV}$ and $0.2\,\textrm{GeV} <P_T <0.4\,\textrm{GeV}$, our model calculation predicts rather flat curves, well agreeing with the data.
We would like to point out that the curve for the range $0.2\,\textrm{GeV} <P_T <0.4\,\textrm{GeV}$ is similar to the calculation on the $g^\perp D_1$ term (the solid line in Fig.~4 of Ref.~\cite{Aghasyan:2011zz})
based on the models from Refs.~\cite{Gamberg:2007wm,Bacchetta:2007wc}.
For higher $P_T$ ranges, deviation between our calculation and the data is found in the larger $x$ region.

In Fig.~\ref{aluvpt}, we display the same asymmetry from our calculation, but as a function of the $\pi^0$ transverse momentum $P_T$. Here, the four panels in Fig.~\ref{aluvpt} correspond to the asymmetry integrated over four different $x$ ranges.
Agreement between the theoretical calculation and the data is found in the lower $P_T$ region ($P_T<0.5 \,\textrm{GeV}$) for $x<0.4$.
More specifically, our theoretical curves increase with increasing $P_T$ in the whole $P_T$ region within the figure, while the experimental data increase with increasing $P_T$ then approach a maximum at around $P_T \approx 0.4 \,\textrm{GeV}$, indicating that our prediction overestimates the experimental data in the region $P_T>0.5 \,\textrm{GeV}$.

From the comparison between the theoretical calculation and the CLAS data, we conclude that the $g^\perp D_1$ term can account for the beam SSA in $\pi^0$ production at CLAS in the regions $x$ and $P_T$ are not large ($x<0.4$ and $P_T <0.5$ GeV).
However, our model overestimates the data in the higher $P_T$ and $x$ regions.
This disagreement might be explained by the absence of other contributions in Eq.~(\ref{FLU}) that we neglect in the calculation (such as the  $h_1^\perp {\tilde{E}}$), or by the possibility that the tree-level factorization is not suitable in this region.

\begin{figure}
  \includegraphics[width=\columnwidth]{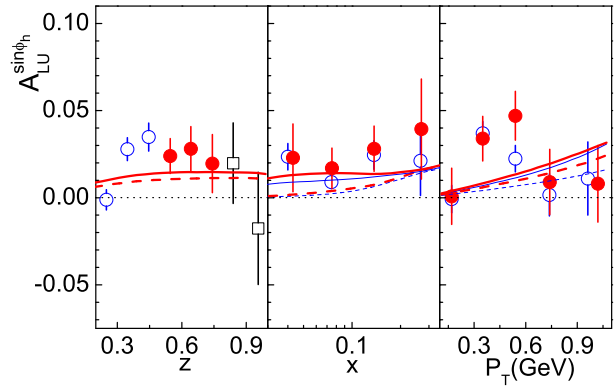}
  \caption{The beam SSA $A_{LU}^{\sin\phi_h}$ for $\pi^0$ production in SIDIS at HERMES vs $z$ (left panel), $x$ (central panel), and $P_T$ (right panel).
  The solid and the dashed lines correspond to the results without and with the kinematical constraints (\ref{constraints}) on $k_T$, respectively.
  The thin and thick lines in the central and right panels correspond to the results for the ranges $0.2<z<0.5$ and $0.5<z<0.8$.
  Data are from Ref.~\cite{hermes07}, with open circles, full circles, and open squares for $0.2<z<0.5$, $0.5<z<0.8$, and $0.8 < z < 1$.
  The error bars represent the statistical uncertainty. }\label{HERMESpi0}
\end{figure}

Furthermore, we also compare our calculation of the $A_{LU}^{\sin\phi_h}$ asymmetry for neutral pion production in SIDIS with the data measured by the HERMES Collaboration~\cite{hermes07}.
The HERMES measurement uses a longitudinally polarized 27.6 GeV positron beam off the hydrogen gas target.
In addition, the following kinematics are applied in the calculation~\cite{hermes07}
\begin{align}
&0.023 < x < 0.4,\,0 < y < 0.85, \,1 \textrm{GeV}^2< Q^2 < 15\, \textrm{GeV}^2, \nonumber\\
& W^2 > 4\, \textrm{GeV}^2,~~
~~ 2\,\textrm{GeV} < P_h < 15\, \textrm{GeV},
\end{align}
where $P_h$ is the energy of the detected final-state $\pi^0$ in the target rest frame.
In the left, central, and right panels of Fig.~\ref{HERMESpi0}, we show the asymmetry vs $z$, $x$, and $P_T$ and compare it with the
HERMES data~\cite{hermes07}.
It is found that the theoretical curves agree with the experimental data within the statistical uncertainty.
The predicted $z$ dependence of the asymmetry is rather flat, which is consistent with the the data in the mid-$z$ region.
The calculation without the constraints in Eq.~(\ref{constraints}) describes the data in the smaller $x$ region better than that with the constraints.

\begin{figure}
  \includegraphics[width=\columnwidth]{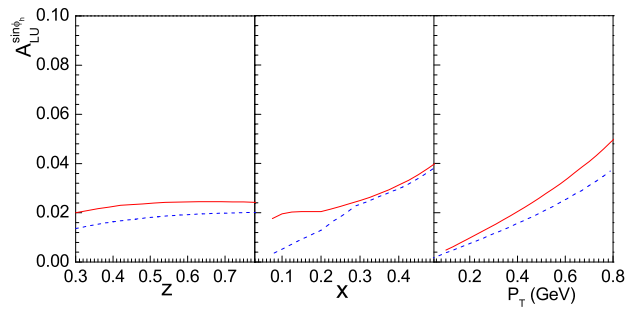}
  \caption{Beam SSA $A_{LU}^{\sin\phi_h}$ of $\pi^0$ at CLAS12 as a function of $z$ (left panel), $x$ (central panel), and $P_T$ (right panel). The solid and the dashed lines correspond to the results without and with the kinematical constraints (\ref{constraints}) on $k_T$, respectively. }\label{clas12}
\end{figure}

Finally, we present the prediction of the asymmetry $A_{LU}^{\sin\phi_h}$ at CLAS12, which can be measured with a $12 \,\textrm{GeV}$ polarized
electron beam. The kinematical cuts for CLAS12 applied in the calculation are~\cite{Avakian:2011zz}
\begin{align}
&0.08 < x < 0.6,~~ 0.2 < y < 0.9, ~~0.3 < z < 0.8,\nonumber\\
&Q^2 > 1\,\textrm{GeV}^2,~~  W^2 > 4 \,\textrm{GeV}^2, ~~0.05\,\textrm{GeV} <P_T< 0.8 \,\textrm{GeV}.
\end{align}
In the left, central, and right panels of Fig.~\ref{clas12}, we plot the $z$,
$x$, and $P_T$ dependencies of the asymmetry, respectively.
Our calculation shows that the beam SSA at CLAS12 is smaller than that at CLAS, but is still sizable.
Our result indicates that there is no obvious $z$ dependence of the asymmetry in $\pi^0$ production; this is understandable since in our approach, the same FFs $D_1^q(z,\bm p_T^2)$ appear in both the numerator and the denominator of the expression for
$A_{LU}^{\sin\phi_h}$.
Therefore, the precision measurement of the $z$ dependence of $A_{LU}^{\sin\phi_h}$ in $\pi^0$ electroproduction at CLAS12 can verify the role of $g^\perp$ in beam SSA.

\section{conclusion}
\label{conclusion}

In this work, we have performed the calculation of the $T$-odd twist-3 TMD distribution $g^\perp(x,\bm k_T^2)$ for the $u$ and $d$ quarks inside the proton in the spectator model with scalar and axial-vector diquarks.
The difference between the isoscalar ($ud$-like) and isovector ($uu$-like) spectators for the axial-vector diquark is considered in the calculation.
We make use of the single-gluon exchange between the struck quark and the spectator to generate the $T$-odd structure.
To obtain a finite result, we choose the dipolar form factor for the nucleon-quark-diquark coupling.
We find that $g^{\perp u}$ and $g^{\perp d}$ have different $x$ and $k_T$ dependencies. First, $g^{\perp u} $ is positive for all $x$ and $k_T$ regions, while $g^{\perp d}$ is negative in the small $x$ region and turns out to be positive in the region $x>0.15$. Second, $g^{\perp d}$ approaches to zero faster than $g^{\perp u}$ with increasing $k_T$.

Using the model results, we analyze the beam SSA $A_{LU}^{\sin\phi_h}$ in semi-inclusive pion electroproduction.
We apply the Wandzura-Wilczek approximation for the FFs; that is, we ignore the contributions from $\tilde{G^\perp}$ and $\tilde{E}$.
Furthermore, we consider the specific case of $\pi^0$ production, in which the $e H_1^\perp$ term should give vanishing contribution because the favored and the unfavored Collins functions have similar sizes
but opposite signs.
Thus, we can perform the phenomenological analysis on the asymmetry $A_{LU}^{\sin\phi_h}$ in $\pi^0$ production at CLAS and HERMES just based on the $g^\perp D_1$ term safely.
The comparison between our theoretical calculation and the data indicates that the $g^\perp D_1$ term can account for the beam SSA in $\pi^0$ production measured by the CLAS Collaboration in the region $x<0.4$ and $P_T<0.5 \,\textrm{GeV}$, where our theoretical curves describe the data fairly well.
In addition, our calculated asymmetry at the HERMES kinematic region is consistent with the HERMES measurements after the error bars of the data are considered.
Our study suggests that the $T$-odd twist-3 distribution $g^\perp$ plays
an important role in the beam SSA in SIDIS, especially in the case of neutral pion production.

\section*{Acknowledgements}
This work is partially supported by National Natural Science
Foundation of China with Grant No.~11005018,
by SRF for ROCS from SEM, and by the Teaching and Research Foundation for
Outstanding Young Faculty of Southeast University.

\end{document}